\documentstyle[12pt]{article}
\vspace*{.75in}
\begin{document}
\baselineskip=22pt plus 0.2pt minus 0.2pt 

\lineskip=22pt plus 0.2pt minus 0.2pt
\begin{center}
 \Large
Quantum Fluctuations and Irreversibility\footnote{To be published in
``The Physical Origins of Time Asymmetry", Eds., J. Halliwell, J.
P\'erez--Mercader and W. Zurek, Cambridge University Press (1993).}.\\

\vspace*{0.35in}

\large

J.\ P\'erez--Mercader\footnote{Also at: Instituto de Matem\'aticas y
F\'{\i}sica  Fundamental, C.S.I.C., Serrano 119--123,
28006 Madrid and the Theoretical Division, Los Alamos
National Laboratory, Los Alamos, New Mexico 87545.}\\

Laboratorio de Astrof\'{\i}sica Espacial y F\'{\i}sica
Fundamental\\
Apartado 50727\\
28080 Madrid
 \\

\normalsize

\vspace{.5in}
February 1, 1993\\
\vspace{.5in}
\vspace{.5in}
ABSTRACT
\end{center}

We study how the effects of quantum corrections lead to
notions of irreversibility and clustering in quantum field theory.
In particular, we consider the virtual ``charge" distribution
generated by quantum corrections and adopt for it a statistical
interpretation. Then, this virtual charge is shown to ($a$)
describe a system where the equilibrium state is at its classical
limit ($\hbar \rightarrow 0$), ($b$) give rise to spatial
diffusion of the virtual cloud that decays as the classical limit
is approached and  ($c$) lead to a scenario where clustering takes
place due to quantum dynamics, and a natural transition from a
``fractal" to a homogeneous regime occurs as distances
increase.

\vspace{1.5in}

\pagebreak

In quantum field theory,  the dynamics stems from  both the
interactions  among the fields  and  the  quantum fluctuations to
which they are subject.  Quantum fluctuations are perhaps the most
fundamental feature of quantum field theory, affecting fields,
their sources and the vacuum in which they evolve.

Induced by  the fact  that  quantum fluctuations  are to a certain
extent  random  $^{\cite{bell}}$,  we will study  how this affects
irreversibility.  Namely,  since  $true$  randomness impairs one's
ability  to  carry out  an exact reconstruction of  the past
history of  the system, it  is  legitimate  to   expect  some
irreversibility at the  microscopic (quantum field) level because
of the presence of quantum fluctuations in this domain.

One possible way to study this irreversibility is to introduce some
fundamental statistical notions into the realm of the quantum field
theory. For example, one can consider the effect that quantum
fluctuations have on the ``charge" density\footnote{We enclose the
word charge in quotes because
 we refer to it in a generic sense, that is, we are not
specifically referring to electric charge. We have in mind the
charge for the source of the Poisson equation satisfied by the
effective interaction energy.} of the Poisson equation satisfied
by the quantum corrected potentials. The quantum fluctuations
affect the interaction energy, and this induces a modified
``charge" density; this  is how the Uehling potential appears in
QED. The statistical notions come in when we interpret the induced
charge density as a probability density, an action that we are
justified in taking due to the mathematical properties of the
charge density induced by quantum--fluctuations. We associate with
the full interaction energy a ``charge" density which (as is done
in cosmology and  astrophysics$^{\cite{layzer}}$) we interpret as
a probability density characterizing the spatial distribution of
virtual charges surrounding the original charges. By doing this,
we are making use of the rich $conceptual$ depth of quantum field
theory and explicitly taking into account that, in the computation
of the quantum corrections at some scale, we are actually
including not only simple two body processes with a fixed and
precise impact parameter of the order of that scale, but the
effects of many body interactions with impact parameter smaller
than the distance scale that we are probing. Because of this,
divergences appear in the theory which renormalization removes and
re--interprets. The physical potential that one measures is thus
subject to fluctuations, and it is impossible to specify the
$exact$ dependence of the potential on the individual components
of the virtual ``charge" density. We may, however, ask about the
probability of occurrence of some particular configuration or
similar questions of a statistical nature. This goal is achieved by
adopting for the virtual ``charge" charge density a statistical
interpretation in term of a probability distribution whose density
is the quantum corrected ``charge" density. By studying the
$effective$ (in the RG $^{\cite{gellmannlow}}$ sense) form of the
interaction energy, one acquires information on how the properties
of the virtual cloud change with scale and, through the statistical
interpretation, gain insight on how changes of scale affect the
dynamics of the cloud in aspects relating to the approach to
equilibrium or even structure and form generation properties.

We can obtain the effective interaction energy by solving its RGE.
The solution to this equation$^{\cite{rothe}}$ yields the
interaction energy as a function of the scale at which one probes
the system, and contains the modifications due to the presence of
quantum fluctuations. Unfortunately, the corrections that are
included to $all$ orders of perturbation theory are only the
leading logarithms; but, as is well known, even with this
limitation, many interesting physical consequences can still be
extracted. Since energy has canonical dimension of inverse time
and no anomalous dimension, the RGE has the solution,

\begin{equation}
V(\lambda r_0,g_0,a)=\lambda ^{-1}V(r_0,\bar{ g}_0(\lambda ),a)
\label{1}
\end{equation}

\noindent
Here $V$ is the interaction energy, $r_0$ is the distance between
the interacting sources, and $a$ is a reference distance. The
quantity  $\bar{g}_0(\lambda )$ is the effective coupling and
$\lambda$ the scale parameter; $\bar{g}_0(\lambda )$ satisfies the
RGE $\lambda \partial \bar{g}_0 / \partial \lambda = - \beta
(\bar{g}_0)$.

In general, for a massless  mediating  field (such as photons or
gluons), the effective  interaction energy  for two point
particles separated by a distance $a$ is given by

\begin{equation}
V(a,g_0,a)=C\frac{g_0^2}{4 \pi a}
\label{2}
\end{equation}

\noindent

Computing to 1--loop order, where $\beta=\beta_0 g_0^3$,  choosing
$\lambda =r/a$, and taking  $\sigma$  to  be  a
constant\footnote{Strictly speaking, $\sigma$ is a function of the
distance at which the system is probed. This can be made explicit
by computing it in a mass dependent substraction
procedure$^{\cite{gandp}}$. However, it will be sufficient for our
present purposes to take it to be a constant.} (for $r$ close to
$a$)

\begin{equation}
V(r,g_0,a) =C\frac{g_0^2}{4 \pi}a^{-\sigma} r^{-1+\sigma}
\label{8/1}
\end{equation}

\noindent
$\sigma$ is related to the $\beta$--function for the coupling $g_0$, and to
one--loop
is given by $\sigma=- 2 \beta_0 g_0^2$. It  has three essential
properties: ($i$) it is proportional to $\hbar$, ($ii$) it vanishes
smoothly in the limit of zero coupling and ($iii$) as defined, it is
a function of the momenta which essentially counts  the number of
degrees of freedom excited from the  vacuum  at  distances less
than $r_0$.

For  the case of QED one recognizes  here  the first term  in the
short  distance  expansion  of  the   famous  Uehling  potential; in
QCD one recognizes the expression for the interquark potential.

According to potential theory$^{\cite{doob}}$, Equation
(\ref{8/1}) satisfies a Poisson equation whose right hand side is
proportional to the ``charge" density dictated by quantum
corrections. This   charge  density describes how the phenomenon
of vacuum polarization  has modified the  vacuum  and given rise
to  the formation  of  virtual pairs that,  as is well known,
affect the strength and properties of the interaction in a
substantial way.

However, we have no information as to the actual distribution of
these pairs, but by interpreting the vacuum  polarization charge density
as a ``probability density", we can derive information on the
virtual cloud and its physics as a many body (statistical) system.
{}From the  mathematical point of view this is possible thanks to the
intimate relationship$^{\cite{doob}}$ that exists between potential
theory and probability theory. The relationship between the
potential and the charge density (away from $r=0$) is through the
Poisson equation $\nabla ^2\phi =+4\pi \rho (r)$, which for our
isotropic potential gives

\begin{equation}
\rho (r)=A' \ \ r^{-3+\sigma}
\label{10-[1]}
\end{equation}

\noindent
where $A'$ is a constant. We only need to require from the
density $\rho (r)$ that it be a positive and integrable
function on its support in order that it can actually be interpreted
as a probability density (see below).

We  now examine some  properties of Eq.(\ref{10-[1]}).  First of all  we
see that $\rho (r)$ is the solution to the functional equation

\begin{equation}
\rho (\lambda r)=\lambda ^\beta \rho (r)
\label{functional}
\end{equation}

\noindent
with $\beta= -3+\sigma$;  this implies right away that $\rho(r)$ describes
a fractal distribution  of  charge embedded in  3--dimensional
configuration space, and with a Hausdorff (or fractal) dimension
given by $d_f=+\sigma$ . This is not surprising, since $\sigma$ has its
origin in the deviation from $canonical$ scaling due to the quantum
fluctuations. Furthermore,  since $\sigma$ changes as we change the size
of the domain  on which  we probe the  virtual cloud,  it turns out then
that the Hausdorff dimension also changes and we are  then dealing
with a multifractal. For QCD and Quantum Gravity in its
asymptotically free regime, $\sigma$ is positive. In QED and other
non--asymptotically free theories, $\sigma$ is negative. When the
Hausdorff dimension is positive, one sees intuitively that there is
a natural tendency to suppress these fluctuations since, contrary
to what happens in the case of a negative  Hausdorff  dimension,
the configuration space ``cannot accommodate them":  it is not big
enough!

To reveal some of the physical consequences of the randomness associated
with these quantum fluctuations, we will study a few of the
features of the ``statistical mechanics"  of the density in
Eq.(\ref{10-[1]}). In particular, we will look  at  the entropy
associated  with  $\rho$ and also  will give a glance  at  the
nature  of  the  stochastic  processes  that  it supports.

Requiring normalizability of the probability density, we get

\begin{equation}
\rho=A r^{-3+\sigma}
\label{density}
\end{equation}

\noindent
with

\begin{equation}
A={\sigma  \over {4\pi }}R_0^{-\sigma }
\end{equation}

\noindent
for $\sigma > 0$, and

\begin{equation}
A=-{\sigma  \over {4\pi }}r_0^{-\sigma }
\end{equation}

\noindent
when $\sigma < 0$. Here $R_0$ denotes an IR cutoff, necessary in the
case of positive $\sigma$, in order to ensure the finiteness of the
probability distribution. For non--asymptotically free theories, $r_0$
is an UV cutoff which becomes necessary for the same reasons. The IR
cutoff can be identified, e. g., with a typical hadronic size and $r_0$
with Planck's length. The resulting probability density is of the
Pareto type$^{\cite{schroeder}}$; this is a natural consequence of
the renormalization group origin for $\rho (r)$, which is ultimately
responsible for the functional equation in (\ref{functional}) and
the associated scaling. Scaling leads to Levy--type distributions,
which turn Pareto in some limit. Notice also that both densities go
to zero in the classical ($\hbar \rightarrow 0$) limit.

To gain information about the ``equilibrium configurations" supported
by these distributions, we construct and compute an entropy--like
quantity. We will assume that, as in
thermodynamics, the state of equilibrium
corresponds to the maximum for the entropy. In other words, we will
assume that it reveals a preferred ``direction"  for stability in
the evolution of the quantum field system. It is possible to
discuss two types of entropies: a ``coarse grained" entropy based
on the probability distribution, and a ``fine grained" entropy or
``differential" entropy\footnote{In the same sense as in information
theory$^{\cite{coverandthomas}}$.}, based on the probability
density $\rho (r)$. We will discuss only the latter, which is
defined through

\begin{equation}
S=-k\int_{}^{} {d^3}\vec r\rho (r)\log \left[ {C\rho (r)} \right]
\end{equation}

\noindent
with the integral extended over the full support of the variable $r$.
The constant $C$ needs to be introduced because $\rho (r)$ is a
dimensional quantity, and we do not have the equivalent of a Nernst
theorem to set a reference valid for $all$ physical systems. This
constant, as in thermodynamics,$^{\cite{fermi}}$ fixes the minimum entropy of
the system.
We will set Boltzmann's constant $k$ equal to one.

The differential entropy for the case when $\sigma$ is positive gives

\begin{equation}
S^{\left( {\sigma >0} \right)}=1-{3 \over \sigma }-\log {\sigma  \over
{4\pi }}C+3 \log R_0
\end{equation}

\noindent
and for negative $\sigma$ one gets,

\begin{equation}
S^{\left( {\sigma < 0} \right)}=1-{3 \over \sigma }-\log {-\sigma  \over
{4\pi }}C+3 \log r_0
\end{equation}

{}From these two expressions we see that the ``entropy constant", $C$, may
be related in an interesting and useful way to the ``extreme" volume (the
volume at the cutoff) of the quantum system. Such a choice has the
advantage of eliminating the dependence of the entropy on the cutoff. For
asymptotically free theories, if we choose $C$ proportional to the
maximum volume, i.e., proportional to the volume of the IR-cutoff to
the cube, then the dependence on the cutoff and extreme system size
disappears from the expression for the entropy. The equivalent
statement is also true for the case of non--asymptotically free
theories, where the ``Nernst--volume" corresponds to the minimum
volume that can be physically reachable.

These entropies are shown in Figures 1 and 2.

We see that in asymptotically free theories, the equilibrium state
corresponds to the $largest$ possible value\footnote{For positive
$\sigma$, $S^{\left( {\sigma > 0} \right)}$ has a maximum at
$\sigma = 3$; unfortunately, this value lies outside the range
where we can reasonably trust the approximations made here.} of
$\sigma$. Now, since matter fluctuations contribute negatively to
$\sigma$, the most stable state corresponds to a configuration (or
system size) where the matter states do not contribute. Because of
the decoupling theorem$^{\cite{decoupling}}$, this occurs for the
{\it maximum--possible size} of the system: that is for its
$classical$ limit!

On the other hand, when the theory is non--asymptotically free, the
maximum of the entropy happens when $\sigma$ goes to 0 from the left.
This means either the classical limit, in the sense that $\hbar
\rightarrow 0$, or that one probes the system at distances {\it so
large} that no quantum fluctuation contributes to $\sigma$.

Thus, for both, asymptotically--free and
non--asymptotically--free  theories one sees that {\it the maximum of
the entropy is attained at the classical limit}.

How does this transition to the equilibrium state take place?
Since diffusion is a mechanism for the transition to the
equilibrium state for a system where randomness is at work, we
will examine diffusion in quantum field theory, and check if the
emerging picture is consistent with what we have obtained from the
entropy. In
quantum field theory, the  fractal nature of the probability
density leads us to expect some form of fractal diffusion or, more
precisely, diffusion through fractal brownian motion. Because of
this ``fractality" we $expect$ the diffusion to be non--space
filling, and to lead to either clustering ($\sigma >0)$ or
screening when the system relaxes; it is clear that this
behavior ought to depend on the sign of $\sigma$. In what follows
we will examine some of the properties of the coefficient of
diffusion derived from $\rho (r)$; then we will examine
the correlation integral in some specific cases.

The virtual charge density cloud described  by the probability density
Equation (\ref{density}) will ``diffuse" in a manner
similar$^{\cite{chandrasekar}}$ to what occurs in  Brownian
motion. We expect that the components of the virtual  cloud
scatter  randomly  among   themselves;  the collision probability
density is $\rho(r)$; the virtual charges execute random walks
controlled by $\rho(r)$; the net effect of these processes is that
the virtual cloud diffuses outside of the ball  of radius $R$.

As is well known$^{\cite{chandrasekar}}$ the probability that a
``particle"  executing a random walk, with each individual
step governed by a probability density $\rho(r)$, be after
$N$--steps at a position between $\vec r$ and $\vec r + d\vec r$,
 is given by

\begin{equation}
W(r)d^3 \vec r=\left[ {4\pi Dt} \right]^{-3/2}\exp \left[ {-\left|
{\vec r} \right|^2/\left( {4Dt} \right)} \right]d^3\vec r
\label{17/1}
\end{equation}

\noindent
where $D$ is the coefficient of diffusion,

$$
D={n \over 6}\left\langle {\vec r^2} \right\rangle \,\, ,
$$

\noindent
$n$ is the number  of  steps per unit time and $N=nt$ is the total
number of  steps.  $\left\langle {\vec r^2} \right\rangle$
is the second moment  of the probability  density $\rho (r)$. The function
$W(r)$ in
Eq. (\ref{17/1}) satisfies the 3--dimensional diffusion equation

$$
{{\partial W} \over {\partial t}}=D\nabla ^2W
$$

\noindent
with  boundary  condition that  $\left. W \right|_s=0$
 at  an  absorbing  medium, or vanishing normal derivative of $W$ at a
reflecting surface: $\left. {{{\partial W} \over {\partial n}}}
\right|_s=0$. How far  and  how  fast  a  disturbance  propagates
depends  on  the coefficient of  diffusion:  for large $D$ one has large
amounts of diffusion, and viceversa. When $D=0$, there is no diffusion.

     In addition to the the Markovian nature  of the
process, there are two assumptions  involved  in  the  derivation  of
(\ref{17/1}) which we  must bear in  mind:  ($a$)  the number of steps $N$
is very large and ($b$)  in the computation of the characteristic
function, it is assumed that $kr<<1$. These two assumptions mean
in  our case  that  the time  intervals that we consider  are
``long"  compared with  virtual time  intervals and that the size
of the region in  which we expect the  diffusion to take place,  is
``large"  compared to the Compton wavelength about which we  are
computing the coefficient $\sigma$.  Both  of these restrictions
are amply met by the validity of our approximations.

In order to compute the coefficient of diffusion on needs to obtain
the second moment of the probability density. From a
radial distance $s_0$ to a radial distance $s > s_0$, the second moment of
$\rho (r)$ is

\begin{equation}
\left\langle {\vec r^2} \right\rangle=\frac{4 \pi}
{2+\sigma}  A
 \left(
s^{2 + \sigma} -s_0^{2+\sigma}
\right)
\end{equation}

\noindent
and therefore, in some cases, the coefficient of diffusion
$diverges$ (naively) as $s$ goes to infinity. This again is not
surprising,  since our probability distribution is  the limit of a
Levy--type distribution, and thus  its moments are divergent.

As previously, we classify the situation depending on whether
we have an asymptotically free theory or not. When $\sigma > 0$, we
can take $s_0 =0$ and set $s=R_0$, the IR cutoff. The coefficient of
difussion is given by

\begin{equation}
D^{(\sigma > 0)}=\frac{n}{6} \frac{\sigma}{2+\sigma} R_0^2.
\label{D+}
\end{equation}

\noindent
In non--asymptotically free theories, we take $s_0=r_0$ (the
UV--cutoff) and leave $s$ unspecified. Then,

\begin{equation}
D^{(\sigma < 0)}=\frac{n}{6} \frac{-\sigma}{2+\sigma} s^2
\left[
\left(
\frac{s}{r_0}
\right)^{\sigma}
-
\left(
\frac{r_0}{s}
\right)^2
\right].
\label{D-}
\end{equation}

\noindent
The expression for $D^{(\sigma >0)}$ reflects the divergence in the
second moment of Levy--type distributions. However, quantum
corrections again play an important r\^ole here: physics limits
the size of the system ($R_0$ cannot go to infinity) and in the
classical limit, when $\sigma \rightarrow 0$, difussion $stops$.
Before this happens, i.e., when the typical energies are larger
than the inverse Compton wavelength of the IR--cutoff, the system
will tend to diffuse itself into clusters or well differentiated
pieces. This last statement being a consequence of the positive
fractal (Hausdorff) dimension associated with $\sigma >0$, as was
mentioned above.

For $\sigma <0$ the situation is qualitatively different. We can
distinguish two different regimes within non--asymptotically free
theories, according as to whether $0>\sigma >-2$ or $-2 > \sigma$. In
the former case, $\left\langle {\vec r}^2 \right\rangle$ diverges as
diffusion takes place into larger distances, and is only shut--off by
the vanishing of $\sigma$ with distance, that is to say, by the
transition into the classical regime. When $\sigma < -2$, the
coefficient of diffusion goes to zero as $s$ increases; this means
that $new$ states stationary under diffusion can be created and
supported at these very deep values of $\sigma$. When $\sigma > -2$
the diffusion coefficient diverges for large $s$ and quantum
corrections will shut the diffusion process off by eventually driving
$\sigma$ to zero; what happens is that  as $s$ increases,  the diffusion
coefficient at  first increases,  to then decrease and cut itself off to
zero when $\sigma$  goes to zero  or  when the classical regime ($\hbar
\rightarrow 0$)  is reached.

We have seen how quantum corrections lead into diffusion, and in some
cases, clustered states of the quantum field system. To get more
information about clustering and other structure formation
properties, it is convenient to look at the correlation
function and integral; from an analysis of these objects one can
get a better feeling of the structures$^{\cite{darcy}}$ supported by
the quantum field theory. The correlation integral, $C(s)$, gives
information  on  how many correlated pairs there are whose
separation is $less$ than $s$. This integral is computed from  the
correlation  function $\xi (\vec r)$ by  the following
formula$^{\cite{grassberger}, \cite{layzer}}$

$$
C(s)=\int\limits_{V(s)} {d^3}\vec r\left[ {1+\xi (\vec r)} \right]
$$

\noindent
where  the region of integration extends over the  volume  $V(s)$ of
radius $s$.  The  correlation function $\xi (\vec r)$ is  related  to
$\delta(\vec k)$,  the Fourier transform of the density $\rho (\vec
r)$, via the following Fourier transform

$$
1+\xi (\vec r)=\left( {2\pi } \right)^3\int {d^3}\vec k e^{-i \vec
k\cdot \vec r}\left| {\delta (\vec k)} \right|^2
$$

\noindent
where

$$
\delta (\vec k)=\int {d^3}\vec re^{+i\vec k\cdot \vec r}\rho
(\vec r)
$$

\noindent
The quantity $\left| {\delta (\vec k)} \right|^2$ is called the
power spectrum, and  measures  the  mean  number  of  correlated
neighbors  in $excess$ of random over a distance of order $1/k$;
because of this,  the  quantity  $1+\xi (\vec r)$  (once
appropriately  normalized) gives information on whether there is
``clustering"  ( $> 1$), ``voids" ($<1$) or  a  perfectly  random
distribution  ($=1$).  This is a straightforward program that
can be carried out for $\rho (r)$. One obtains the
following expressions for $1+\xi (\vec r)$ and $C(r)$

$$
1+\xi (r)=D(\sigma) \cdot r^{-3+\sigma}
$$

$$
C(r)=\frac{2 \pi}{\sigma} \cdot D(\sigma) \cdot r^{2 \sigma}
$$

\noindent
where the coefficient $D(\sigma)$ is

$$
D(\sigma)=(4\pi A)^2\cdot 4\pi \cdot \left| {B(\sigma )}
\right|^2\cdot \Gamma (2-2\sigma )\cos {\pi  \over 2}(2\sigma -1)
$$

\noindent
and

$$
B(\sigma)=\Gamma (\sigma -1)\cos {\pi  \over 2}(2-\sigma).
$$

$D(\sigma)$ is shown in Figure 3, and in Figure 4 we show
$D(\sigma)/\sigma$, which are important in determining the behaviors
of $1+\xi (r)$ and the correlation integral.

{}From Figure 3 we see that for positive $\sigma$ and between zero
and 3/2, the quantum field system leads to clustering. For
$\sigma$ between 3/2 and 5/2 there is a significant qualitative
change into a scenario of deep anticlustering (``voids") with a
tendency for the system to behave near homogeneity when $\sigma
\approx 2$. When $\sigma$ increases past 5/2, we enter a new
clustering domain. The rest can be ``read off" in the same way
from the figure. The plot of Figure 4 simply confirms the one for
$D(\sigma)$, but for the coefficient of the correlation integral.
We also point out that there is a ``quantization" of structures, a
kind of periodicity, that takes place as $\sigma$ changes and goes
through some critical values.

We end by summarizing our results. Quantum field theory is a many
body system par excellence. Intrinsic to it is the randomness
associated with quantum fluctuations. The application of a
statistical interpretation to one of the most primitive concepts of
field theory, that of the induced charge, has interesting and
apparently deep consequences which stem from its scaling
properties. In particular, fractal behavior in a form related to
the Pareto--Levy form of the charge density, leads to systems where
the maximum entropy is associated with the classical limit of the
quantum system. In other words, the classical regime is the most
stable, and the one preferred by the system.

Like in other many body systems, diffusion is a familiar mechanism
for the approach to equilibrium (although perhaps not the only
one). In concordance to what follows from studying the entropy, the
virtual cloud diffuses with a coefficient of diffusion proportional
to $\hbar$, and thus diffusion of the cloud stops in the classical
limit.

{}From an analysis of the correlation properties of the virtual cloud
system, one learns that the system supports the formation
of structures where there is a sharp transition between phases of
anticlustering (void formation), clustering and some interspersed
quasi--homogeneity. These depend quite clearly on the number of
degrees of freedom excited from the vacuum into the system; in
other words, they depend on the distance at which the system is
probed.

In very general terms, we see that irreversibility is contained in
quantum field systems, and that this irreversibility is
encountered as the system size grows. The quantum field system
tends to ``relax" into a classical system or into clustered
structures, depending on the scale of the system and the nature of
the interactions. The ideas developed here may also find
application in the study of intermittency, turbulence, phase
transitions, multiparticle physics and other complex phenomena in
quantum field theory and the early universe.

\noindent

\vspace{.25in}
\Large \bf{Acknowledgements.}
\vspace{.25in}

\normalsize
\noindent
This workshop was financed by Fundaci\'on Banco de Bilbao Vizcaya
and by NATO. The Fundaci\'on also provided us with logistic,
material and public support. I wish to thank Don Jos\'e Angel
S\'anchez Asia\'{\i}n, Do\~na Maria Luisa Oyarz\'abal and Don Jos\'e
Ignacio Oyarz\'abal for their enthusiastic support. I would also
like to thank my co--organizers, Jonathan Halliwell and Wojciech
Zurek for the pleasure and fun in working together. I have
benefitted from discussions with O. Bertolami, R. Blankebeckler,
A. Carro, T. Goldman, A. Gonz\'alez--Arroyo, S. Habib, J. Hartle, R.
Laflamme, S. Lloyd, C. Morales, M. M. Nieto, E. Trillas,
G. Veneziano and G. West.


\begin{thebibliography}{99}

\bibitem{bell} Bell, J. S., ., {\sl Speakeable and unspeakeable in
quantum mechanics}, Cambridge University Press, Cambridge, 1991.

\bibitem{layzer} See for example, Layzer, D., in {\sl Galaxies and
the Universe}, edited by Sandage, A., Sandage, M.  and Kristian,
J.,  University of Chicago Press, Chicago, 1975.

\bibitem{gellmannlow} Gell--Mann, M. and Low, F. E., {\sl Phys. Rev.
95 (1954) 1300--1312}. For a beautiful presentation full of
insight, see West, G. in {\sl Particle Physics: A Los Alamos
Primer}, edited by Cooper, N. and West, G., Cambridge University
Press, Cambridge, 1988.

\bibitem{rothe} Rothe, H. J., {\sl Lattice Gauge Theories}, World
Scientific, Singapore, 1992.

\bibitem{gandp} Georgi, H. and Politzer, D., {\sl Phys. Rev. D14
(1974) 451}.

\bibitem{doob} Doob, J., ., {\sl Classical Potential Theory and Its
Probabilistic Counterpart}, Springer--Verlag, New York, 1984.

\bibitem{schroeder} Schroeder, M., {\sl Fractals, Chaos, Power
Laws}, W. H. Freeman and Co., San Francisco, 1991.

\bibitem{coverandthomas} Cover, T. M. and Thomas, J. A., {\sl
Elements of Information Theory}, J. Wiley and Sons, New York, 1991.

\bibitem{fermi} Fermi, E., {\sl Thermodynamics}, Dover Books, New
York, 1956.

\bibitem{decoupling} Appelquist, T. and Carrazone, J., {\sl Phys.
Rev.  D11 (1975) 2896}.

\bibitem{chandrasekar} Chandrasekhar, S., {\sl Rev.  Mod. Phys.15
(1943) 1--89}.

\bibitem{darcy} Thompson, D'Arcy Wentworth, {\sl On Growth and
Form: The Complete Revised Edition}, Dover Books, New York, 1992.

\bibitem{grassberger} Grassberger, P. and Procaccia, I. {\sl
Physica D9 (1983) 189}.

\end{thebibliography}
\end{document}